\documentclass[aps,pre,preprint,superscriptaddress,showpacs,floatfix]{revtex4}
\usepackage{amssymb}
\usepackage{epsfig}
\usepackage{amsmath}
\usepackage{graphicx}
\usepackage{xcolor}
\begin{document}
\title{Explosive synchronization transition in a ring of coupled oscillators}
\date{\today}

\author{Wei Chen}
\affiliation{School of Science, Beijing University of Posts and
Telecommunications, Beijing 100876, China.\\}

\author{Weiqing Liu}
\email{wqliujx@gmail.com} \affiliation{School of Science, Jiangxi
University of Science and Technology, Ganzhou 341000, China.\\}

\author{Yueheng Lan}
\affiliation{School of Science, Beijing University of Posts and
Telecommunications, Beijing 100876, China.\\}

\author{Jinghua Xiao}
\email{jhxiao@bupt.edu.cn} \affiliation{School of Science, Beijing
University of Posts and Telecommunications, Beijing 100876,
China.\\}
\affiliation{State Key Lab of Information Photonics and Optical Communications, Beijing University of Posts and Telecommunications, Beijing 100876, China.\\
}

\begin{abstract}
Explosive synchronization(ES), as one kind of abrupt dynamical
transition in nonlinearly coupled systems, is currently a subject of
great interests. Given a special frequency distribution, a  mixed ES
is observed in a ring of coupled phase oscillators which transit
from partial synchronization to ES with the increment of coupling
strength. The coupling weight is found to control the size of the
hysteresis region where asynchronous and synchronized states
coexist. Theoretical analysis reveals that the transition varies
from the mixed ES, to the ES and then to a continuous one with
increasing coupling weight. Our results are helpful to extend the
understanding of the ES in homogenous networks.

\end{abstract}

\keywords{Explosive synchronization; Coupled oscillators}

\pacs {05.45.Xt, 05.65.+b}

\maketitle

\begin{quotation}
\textbf{Explosive synchronization transitions in large number of
coupled complex network have been a hot topic since it is related to
many inner regimes of collective dynamics and emergency phenomena.
Although explosive synchronization in heterogeneous networks are
commonly observed, the formation of it in homogeneous networks is
still far beyond understanding. A kind of mixed explosive
synchronization is firstly observed in the ring of coupled phase
oscillators with a special frequency distribution where the coupled
oscillators transit from partial synchronization to explosive
synchronization with the increasing coupling strength. We further
report on the effects of coupling weight on controlling the extent
of the hysteretic region of coexistence of the unsynchronized and
synchronized states. Theoretical analysis are presented to reveal
the transition processes from mixed explosive synchronization,
explosive synchronization and then to the continues transition with
the effects of coupling weight. Our results are helpful to
understand the regimes of explosive synchronization in homogenous
networks.}
\end{quotation}

\section{Introduction}
Synchronization, as one of the commonly observed collective behavior
in coupled oscillators, is a hot subject in nonlinear science since
it is related to self-organization in both science and engineering
applications such as neurons, fireflies, or cardiac pacemakers
\cite{1}. Inspired by the research on the small-world or scale-free
networks, various synchronization dynamics in complex networks have
been widely studied \cite{wang}. Generally, two types of transition
from incoherent to coherent state may exist in oscillators coupled
with complex network structure: abrupt percolation and continuous
phase transition. The former one, named as ES, was observed in
scale-free networks of phase oscillators \cite{2,3} and chaotic
oscillators\cite{4}. Kim  et al. \cite{5} explored the ES in the
human brain networks and have tried to explain how it emerges from
unconsciousness in sleep and anesthesia. Skardal et al. \cite{6}
found that ES can be realized by introducing disordered frequency
distribution in synthetic networks. To better disclose the general
mechanism of ES in coupled oscillators, growing attentions have been
paid on the network topology and generic dynamics of nodes. Recent
works \cite{7,3} showed that positive (negative) correlation between
the degrees and the natural frequencies of nodes may lead to ES
(hierarchical synchronization) among oscillators in scale-free
networks, which was experimentally verified in electronic circuits
\cite{8,9}. ES can also be obtained for any given frequency
distribution if the complex networks are constructed with the
frequency disassortative of node degrees\cite{10,11}. Thu et al.
\cite{12} revealed that only when both the degrees and frequencies
of the nodes in a network are disassortative can the ES occur. Leyva
et al. \cite{9} revealed that the  continuous transition may change
to a sharp, discontinuous phase transition in complex network. Zheng
et al. \cite {13} found that frustration may enhance or delay the
explosive transition . Up to now, the investigations on the
mechanism and the ES was mainly focused on network
structure\cite{14,15,16}, dynamics of nodes\cite{17,18,19}, effects
of time-delay \cite{20,21,22,23}, and the coupling
form\cite{11,24,25}. In a heterogeneous scale-free network, the ES
is well understood, due to the positive correlation between degrees
and oscillation frequencies.

 Although ES in heterogeneous network
are commonly observed, the necessary and sufficient condition for
its formation is still unknown. Hou et al.\cite{26} verified that no
ES can be observed by introducing positive correlation between node
degrees and natural frequencies in ER networks. Leyva et al.
\cite{4} observed the ES dynamics in homogeneous Erd\H{o}s-R\'{e}nyi
(ER) networks and all-to-all networks by using frequency dependent
coupling weights. They presented the first precise conditions of ES
in homogeneous networks. However, it is still unclear whether ES can
be observed in a regular ring of coupled oscillators which is
another common and simple network. Rich dynamics exist in a ring of
coupled oscillators such as various kinds of pattern formations
\cite{27} and amplitude death \cite{28}. In our previous work
\cite{29}, we observed abrupt partial synchronization in a ring of
coupled oscillators when the nodes are specially arranged according
to their frequencies. Meng Zhan et al. \cite{30} extended the work
of frequency configuration in a ring of coupled oscillators.
Motivated by these results, it seems natural to generate ES with
appropriate arrangement of nodes based on frequency configuration
and to reveal the generic properties of ES in a ring of coupled
phase oscillators.

The remainder of this paper is organized as follows. In Sec. II, we
introduce the model of coupled phase oscillators and give the main
results. In Sec. III, theoretical results are presented to reveal
the parameter regime for the existence of mixed ES in coupled phase
oscillators. In Sec. IV, the effects of coupling weights on the
synchronization is numerically and theoretically studied.  In Sec.
V, we summarize our studies and point out possible further
development along the line of current study.

\section{Model and results}

Without lack of generality, we consider the most classical and
simple Kuramoto model \cite{31}
\begin{eqnarray}\label{eq1}
\dot{\theta_i}(t)=\omega_{i}+\frac{\epsilon}{k_i+1}\sum_{j=1}^{N}
A_{i,j} \sin(\theta_j-\theta_i),i=1,2...,N,
\end{eqnarray}
where $\theta_i$ is the phase of the $i$th oscillator ($i = 1,2 . .
. ,N$), $\omega_i$ is its associated natural frequency drawn from a
frequency distribution $g(\omega)$($g(\omega)$ follows uniform
distribution in $[0,1]$ for simplicity), $\epsilon$ is the coupling
strength, $A_{i,j}$ is the adjacency matrix with $A_{i,j}=1$ if node
i and node j has connection, otherwise $A_{i,j}=0$. $k_i$ is the
degree of the $i$th oscillator. Here we only focus on the
nearest-neighbor-coupled ring of oscillators (with
$A_{i,(i+1)}=1,A_{(i-1),i}=1,i=2,...,N-1$, and
$A_{N,1}=1,A_{1,N}=1$, otherwise $A_{i,j}=0$).

To measure the transition from incoherence to phase synchronization,
a classical order parameter for the system provided by Eq. \ref{eq1}
is $r(t)=\frac{1}{N}| \sum_{i=1}^{N} e^{j\theta_i(t)}|$, and the
level of synchronization can be measured according to the value of
$R = <r(t)>_T$, with$<...>_T$ denoting time average over a large
time span $T$ and $j=\sqrt{-1}$. When $R\approx 1$, the system
reaches the completely synchronous state. Otherwise, when $R \approx
0$, the oscillator ensemble exhibits incoherence, and oscillators
behave almost independently. With the increment of the coupling
strength $\epsilon$, the coupled system undergoes a phase transition
(second-order or first-order) from  non-synchronous ($R \sim
\frac{1}{\sqrt{N}}$) to synchronous state ($R\approx 1$). In the
following, we will explore the conditions and the parameter regimes
for the occurrence of the ES in a ring of coupled phase oscillators.
\begin{figure}
\includegraphics[width=16cm]{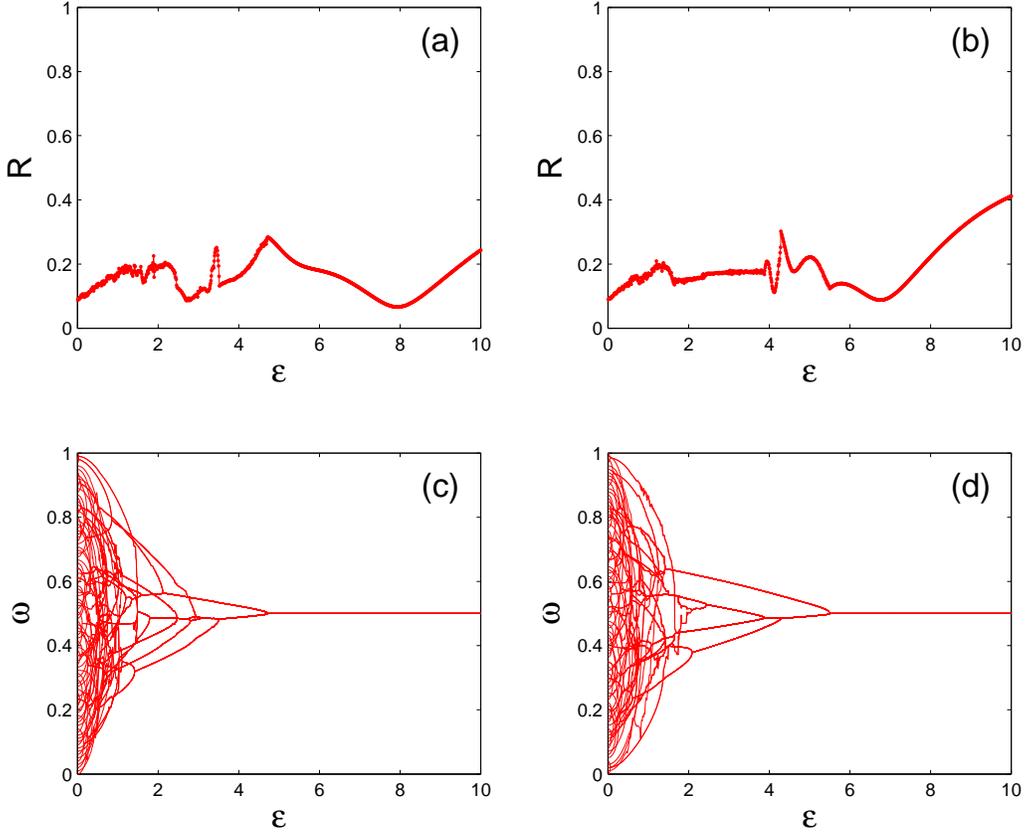}
\caption{(color online) (a)(b) The average frequencies of all nodes
versus coupling strength $\epsilon$ for $N=100$ with two arbitrary
sets of frequencies.(c)(d)  The numerical order parameter $R$ versus
coupling strength $\epsilon$ corresponding to (a)(b),respectively.}
\label{fig1}
\end{figure}
For arbitrarily given frequency configuration of coupled phase
oscillators, No ES dynamics are observed in a ring of coupled phase
oscillators. The average frequencies of the coupled oscillators
versus coupling strength $\epsilon$ are presented in Figs.
\ref{fig1}(a) $\sim$ (b) for two arbitrary sets of random initial
natural frequency configuration. Obviously, the synchronous
transition process of the coupled oscillators is a continuous one
since the final synchronous cluster is built up by combining several
small synchronous clusters. Moreover, the corresponding order
parameters $R$ also exhibit continues transitions as plotted in
Figs. \ref{fig1}(c) $\sim$ (d). What should be mentioned is that the
value of $R$ keeps small even when the oscillators are synchronized.
The reason is that most of the oscillators are phase locked in
anti-phase states in the ring of coupled oscillators. However, based
on our former work \cite{29}, we expect to observe ES in a ring of
coupled phase oscillator when the necessary coupling strength for
synchronization is small with the special frequency arrangement. For
simplicity, we first set the frequency of node $i$ to be
$\omega_i=\frac{i-1}{N-1},i=1,2,...,N$, then rearrange them to a
special configuration with large roughness
$Ro=\frac{1}{N}\sum^N_{=1}(\omega_{i+1}-\omega_i)^2$ \cite{29} as
shown in Fig. \ref{fig2}(a) which is more precisely given by the
following equation.
\begin{eqnarray}\label{eq2}
\left \{
\begin{array}{lll}
\omega_i&=&\frac{N-1+(-1)^i(2i-1)}{2(N-1)},
i\leq\frac{N}{2},\\
\omega_i&=&1-\omega_{N+1-i},i>\frac{N}{2}
\end{array}
,i=1,2...,N.
\right.
\end{eqnarray}
In the above configuration, the oscillators are split into two
groups where one group has relatively large mismatches between pair
of neighboring nodes while the other one has relatively small
mismatches. We will see how an explosive transition is generated in
a system of size $N$ with the given frequency distribution. The
order parameter exhibits two stages of transition when the coupling
strength varies from 0 to 1. In the first stage, the order parameter
is linearly increasing with the increment of coupling strength. In
the second stage, the order parameter sharply transits to around 1
with an associated hysteresis part as shown in Fig. \ref{fig2}(b).
By checking the average frequencies of all nodes versus coupling
strength as shown in Fig. \ref{fig2}(c), we find that a sequence of
nodes with smaller frequency mismatches $\delta
\omega_i=\omega_i-\bar{\omega}$ firstly combines into one
synchronization cluster whose average frequency $\bar{\omega}$ is
0.5. Then all the rest get annexed into the synchronous cluster in a
sudden when the coupling strength is larger than the transition
point. Since the synchronous transition is following a partial
synchronization, we name the whole transition processes as the mixed
ES. The occurrence of the mixed ES is not limited to the special initial natural frequency. When the initial frequencies of nodes before rearrangement are randomly selected, the mixed ES still can be observed if the final frequency distribution is rearranged with the characteristic of having maximal value of the roughness Ro and being strictly symmetric with respect to the average frequency of the whole coupled system. Fig. \ref{fig2}(d)-(f) present one example of random selection of the frequencies where the mixed ES can be observed according to the order parameter R and the average frequency versus coupling strength.

\begin{figure}
\includegraphics[width=16cm]{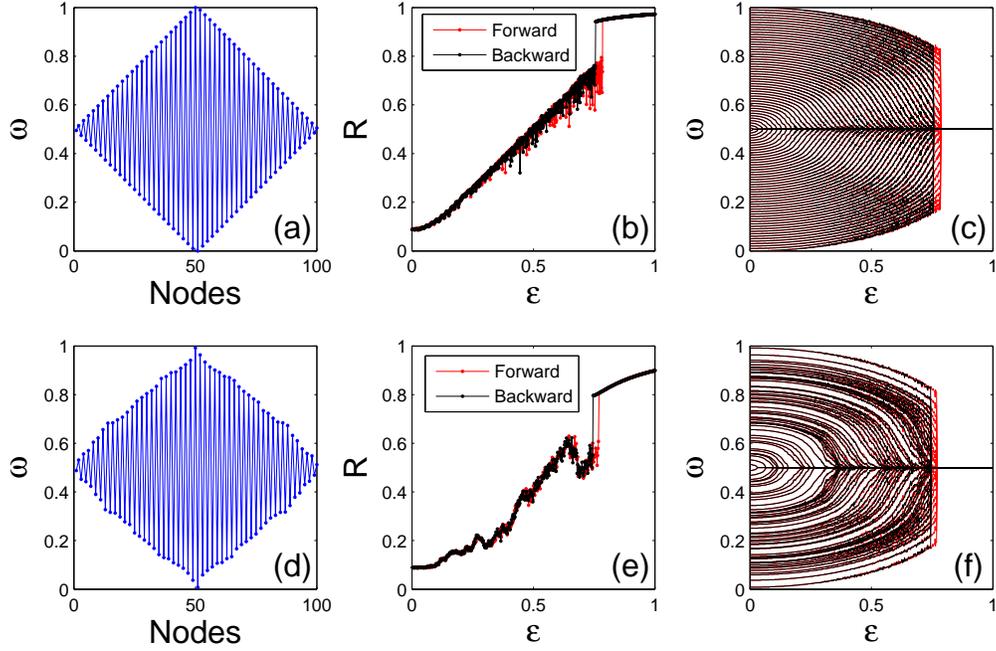}
\caption{(color online) (a) The optimal configuration of the natural
frequency based on Eq. \ref{eq2} of $N=100$ coupled phase oscillators. (b) The order parameter $R$ versus the coupling strength $\epsilon$. The results are recorded by quasi-stationary method where the red lines depict the forward results (increasing $\epsilon$) while the black lines depict the backward results(decreasing $\epsilon$). (c) The average speeds of all nodes versus the coupling strength $\epsilon$ corresponding to (b). The second  row shows the result of the random selection of the frequencies. (d)The rearranged frequency distribution with maximal roughness $Ro$ for one example of randomly set initial frequencies. (e) The order parameter $R$ versus the coupling strength $\epsilon$ for the new set of frequency distribution. (f) The average frequencies of all nodes versus the coupling strength $\epsilon$ corresponding to (e).} \label{fig2}
\end{figure}

\section{Theoretical analysis}
To better understand the dynamics of the ring of coupled oscillators
with the special frequency distribution, we write Eq. \ref{eq1} as
\begin{eqnarray}\label{eq3}
\dot{\theta_i}(t)=\omega_i+\frac{1}{3}\epsilon\Omega_i,
\end{eqnarray}
where
\begin{eqnarray}\label{eq4}
\Omega_i=\sin(\theta_{i+1}-\theta_i)+\sin(\theta_{i-1}-\theta_i).
\end{eqnarray}
According to Eq. \ref{eq2},
$|\omega_{i-1}-\omega_{i+1}|=\frac{2}{N-1}$.
$\omega_{i-1}\approx\omega_{i+1}$ for large $N$. Then
$\sin(\theta_{i+1}-\theta_i)\approx \sin(\theta_{i-1}-\theta_i)$
when nodes $i$,$i-1$,$i+1$ are in a synchronous cluster. Therefore,
\begin{eqnarray}\label{eq5}
\Omega_i\approx 2 \sin(\theta_{i+1}-\theta_i).
\end{eqnarray}
When all nodes are in synchronous cluster, according to the natural
character of the sine function ($|\sin x|\leq 1$), the boundary
values of $\Omega_i$ (here defined as $\Omega_{bws}$ for
convenience) satisfies
\begin{eqnarray}\label{eq7}
\Omega_{bws}=2.
\end{eqnarray}

Based on Eq. \ref{eq3}, if node $i$ is in the synchronous cluster,
then the average frequency of the synchronous cluster can be defined
as $\bar{\omega}=<\dot{\theta_i}>_{T}$ (here $\bar{\omega}=0.5$ for
given frequency distribution in Eq. \ref{eq2} ). Eq. \ref{eq3}
becomes
\begin{eqnarray}\label{eq6}
\Omega_i=\frac{3}{\epsilon}(\bar{\omega}-\omega_i).
\end{eqnarray}

Then the natural frequency of node $i$ in the synchronous cluster
with average frequency $\bar{\omega}$ satisfies Eq. \ref{eq8} for
given coupling strength.

\begin{eqnarray}\label{eq8}
|\frac{3}{\epsilon}(\bar{\omega}-\omega_i)|\leq \Omega_{bws}.
\end{eqnarray}
\begin{figure}
\includegraphics[width=8cm]{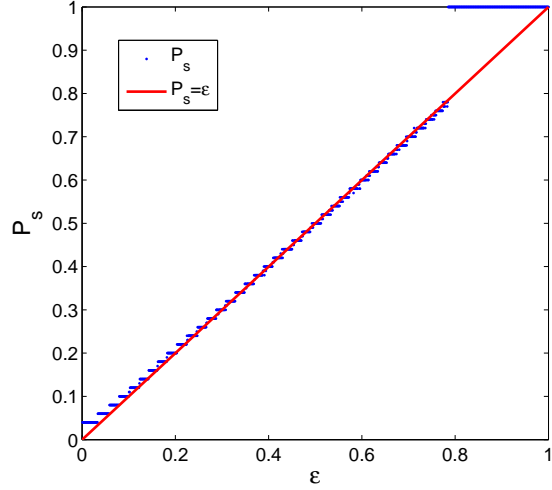}
\caption{(color online) The portion of synchronous nodes versus
coupling strength $\epsilon$ for $N=100$. The red lines depict the linear function $y=a*x$ with the parameter $a=1.000\pm0.0016$ within $95\%$ confidence intervals. }
\label{fig3}
\end{figure}
It should be mentioned that the Eq. \ref{eq8} is fulfilled only when
all nodes are in synchronous clusters. Otherwise, the nodes on the
boundary of synchronous cluster would be pulled out of the cluster
by the neighboring non-synchronous nodes. Therefore, when the
coupled oscillators are in partially synchronous state, the boundary
value of $\Omega_i$ must be smaller than $\Omega_{bws}$ due to the
influence of outer non-synchronous nodes. In order to figure out the
actual boundary value of $\Omega_i$ for partially synchronous state,
we numerically explore the transition process from incoherent to
partial synchronization and try to explore the influence of the
non-synchronous nodes on the synchronous nodes. The results indicate
that the proportion of the synchronized nodes $P_s=\frac{N_s}{N}$ is
equal to $\epsilon$, as shown in Fig. \ref{fig3}, for all $\epsilon$
before the ES. Moreover, the $N_s$ nodes in the synchronous cluster
are those whose natural frequencies are near the value of
$\bar{\omega}(=0.5)$. Define the maximal initial nature frequency of
nodes in the synchronous cluster to be $\omega_{Smax}$, then the
maximal deviation between natural frequency of nodes and the actual
frequency of synchronous cluster can be described as $\delta
\omega_{s}=\omega_{Smax}-\bar{\omega}$. Therefore, according to Eq.
\ref{eq2}, the frequencies of all $N_s$ partially synchronous nodes
are in the range as shown in Eq. \ref{eq9}.
\begin{eqnarray}\label{eq9}
|\omega_i -\bar{\omega}|\leq
\frac{\delta\omega_s}{2}=\frac{\epsilon}{2},
\end{eqnarray}
When the coupled system is in partial synchronization, the actual
boundary values of $\Omega_i$  is
\begin{eqnarray}\label{eq10}
|\Omega_{i}|=\frac{3}{\epsilon}|\bar{\omega}-\omega_i|\leq
\frac{3}{\epsilon}\times\frac{\epsilon}{2}=1.5
\end{eqnarray}
For convenience, we defined $\Omega_{bps}=1.5$ to be the boundary of
$\Omega_{i}$ for partial synchronous clusters, then the actual
boundary of $\sin(\theta_{i+1}-\theta_{i})$ satisfies
\begin{eqnarray}\label{eq11}
|\sin(\theta_{i+1}-\theta_{i})| \leq \frac{\Omega_{bps}}{2}= 0.75.
\end{eqnarray}

\begin{figure}
\includegraphics[width=16cm]{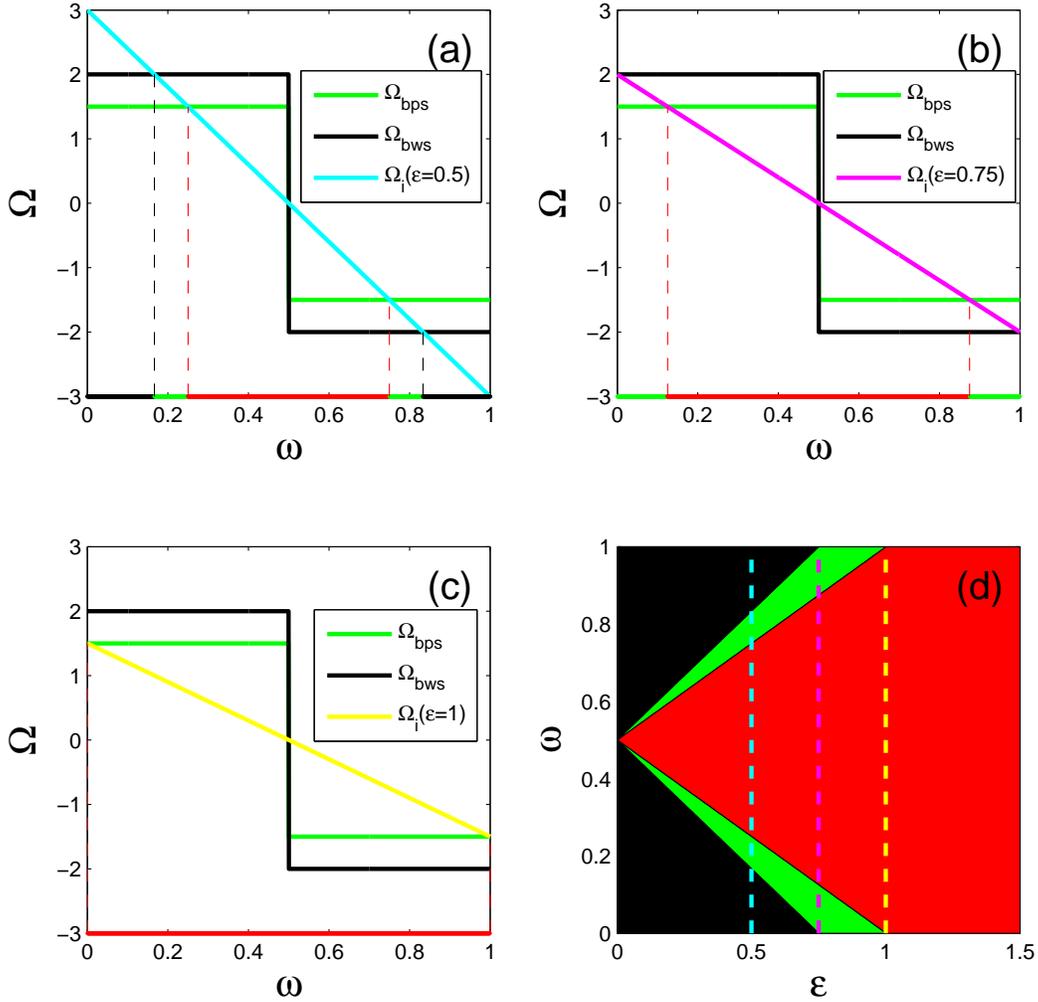}
\caption{(color online) (a)(b)(c) The boundary lines of the full
$\Omega_{bws}$ (black lines) and the partial synchronization
$\Omega_{bps}$ (green lines), and $\Omega_i$ versus natural
frequency $\omega_i$ for $\epsilon=0.5$(cyan lines),
$\epsilon=0.75$(pink lines),$\epsilon=1.0$(yellow lines). (d) Three
types of states as: non-synchronous (black), buffer (green), and
synchronous state(red) in the parameter space of $\omega$ versus
$\epsilon$.} \label{fig4}
\end{figure}

Based on the two boundary values of $\Omega_{bps}$ and $
\Omega_{bws}$, the nodes in the coupled system can be classified
into three types, (1) partially synchronized nodes whose frequencies
$\omega_i$ satisfy $|\Omega_{i}|<\Omega_{bps}$; (2) non-synchronous
nodes whose frequencies $\omega_i$ satisfy
$|\Omega_{i}|>\Omega_{bws}$; (3) buffer nodes whose frequencies
$\omega_i$ satisfy $\Omega_{bps}<|\Omega_{i}|<\Omega_{bws}$. It is
worth mentioning that the fate of the buffer nodes is determined by
the competition between synchronous and non-synchronous group. If
there are non-synchronous nodes in the coupled oscillator system,
the buffer nodes keep away from the synchronous cluster, otherwise,
if the outer non-synchronous nodes disappear, then the buffer nodes
can join the synchronous cluster or stay unsynchronized, which
depends on the initial conditions of those buffer nodes.

To better exhibit the fate of all coupled nodes, we plot the
boundaries of the full synchronization $\Omega_{bws}$(black lines),
partial synchronization $\Omega_{bps}$ (green lines), and the curve
of $\Omega_i$ (blue lines) versus $\omega_i$ in Figs. \ref{fig4}(a)
$\sim$ (c) for given $\epsilon=0.5,0.75,1.0$, respectively.
Obviously, for given $\epsilon$, the slope of the curve $\Omega_i$
is determined
($|\Omega_{i}|=\frac{3}{\epsilon}|\bar{\omega}-\omega_i|$).
Therefore, the fate of the nodes is clearly determined by the
boundary lines and the slope of the curve $\Omega_i$. If the
frequency $\omega_i$ of nodes $i$ satisfies $|\Omega_i|<
\Omega_{bps}$, then the nodes whose frequencies are within the range
between the two intersection points of the boundary lines of
$\Omega_{bps}$ and $\Omega_i$ are partially synchronized (the
frequencies of those nodes are marked with red dots). If
$|\Omega_i|> \Omega_{bws}$, then nodes are in the non-synchronous
state as marked with black dots. If
$\Omega_{bps}<|\Omega_i|<\Omega_{bws}$, then the nodes may go either
to the synchronous or the non-synchronous state depending on the
motion of its neighboring nodes and its initial values (marked with
green dots). The synchronization transition are plotted in the
parameter space of $\omega\sim \epsilon$  as shown in Fig.
\ref{fig4}(d) where the red, green, and black areas marks the
synchronous, the buffer, and the non-synchronous nodes,
respectively. What should be mentioned is that the fate of nodes in
the green area will finally become the non-synchronous state if its
neighboring nodes are in the non-synchronous states (black area for
$\epsilon<0.75$). However, when there are no non-synchronous nodes
for $\epsilon>0.75$, the fate of the buffer nodes can be either in
the synchronous state if the initial condition is near the
synchronous manifold or in the non-synchronous state if the initial
condition is far away from the manifold otherwise, which builds up
the bistable hysteresis area.

According to the analysis above, with the increment of the coupling
strength, the transition from the incoherent to the synchronized
state can be well described by the order parameter $R$. Since the
phase of incoherent oscillators are randomly distributed in
$[0,2\pi]$, they have less contribution to the order parameter $R$.
However, if the oscillators are in the partially synchronous state,
their phase are locked within a range which contribute significantly
to the order parameter $R$. Therefore the order parameter is
approximately calculated as $R\approx R_s$, where $R_s$ is the
contribution of the synchronous nodes. Suppose that there are $N_s$
nodes whose natural frequencies are $\omega_i,i=1,2,..., N_s$ in the
synchronous cluster with average frequency $\bar{\omega}$, then
their phases satisfy  Eq. \ref{eq12}. Moreover, the phase of the
$N_s$ synchronous nodes would be locked to each other (i.e.
$\theta_{i-1}\approx\theta_i$, and
$\sin(\theta_{(i-1)}-\theta_i)\approx \theta_{i-1}-\theta_i$ for
$i=2,3,...,N_s$).
\begin{eqnarray}\label{eq12}
\bar{\omega}=\omega_i+\frac{\epsilon}{3}(\sin(\theta_{i-1}-\theta_i)+\sin(\theta_{i+1}-\theta_i)).
\end{eqnarray}
\begin{figure}
\includegraphics[width=16cm]{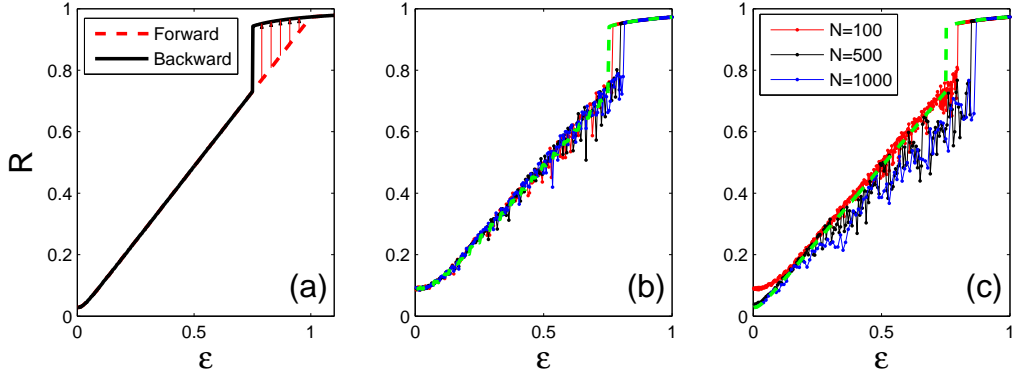}
\caption{(color online) (a)The theoretical results of the order
parameters R versus $\epsilon$. The red (black) lines are results
for the forward (backward) one for N = 1000.  (b)The numerical value
of the order parameters of three arbitrary sets of initial
conditions (red,black and blue) and the theoretical backward results
(green dashed line) for N = 100. (c) The numerical value of the
order parameters for system size N = 100(red dots),N = 500(black
cross),N = 1000(blue circle) and the theoretical backward results
(green dashed line) for N = 1000.}
 \label{fig5}
\end{figure}
we get
\begin{eqnarray}\label{eq13}
\theta_{i-1}-2\theta_i+\theta_{i+1}=2
\arcsin(\frac{3(\bar{\omega}-\omega_i)}{2\epsilon}).
\end{eqnarray}
which determine the value of $\theta_i,i=2,...,N_s$ for arbitrarily
given $\theta_{1}$.
\begin{eqnarray}\label{eq14}
R\approx R_s=<\frac{1}{N}|\sum_{i=1}^{N_s} e^{j\theta_i}|>_T.
\end{eqnarray}
The order parameter $R$ can be determined according to Eq.
\ref{eq14} for the given number $N_s$ of nodes in the synchronous
cluster. It is worth mentioning that $N_s$ are different upon
increasing or decreasing the coupling strength based on the state
distribution in Fig. \ref{fig4} (d).  When decreasing the coupling
strength from $\epsilon=1.0$ to zero, $N_s$ is equal to the number
of the nodes in the red and green area for $\epsilon>0.75$, since
the nodes in green area (buffer nodes) start near the synchronous
manifold and thus get finally synchronized. However, $N_s$ becomes
the number of nodes in the red area when $\epsilon<0.75$ where the
buffer nodes are always in the non-synchronous state. However, if
the coupling strength increases from $\epsilon=0$ to 1, $N_s$
remains equal to the number of nodes in the red area and the buffer
nodes stay away from the synchronous manifolds when non-synchronous
nodes are nearby the buffer nodes. However, when the non-synchronous
nodes nearby the buffer nodes disappear, the buffer nodes may join
into the synchronous cluster without the influence of
non-synchronous nodes. However, when it will jump to synchronous
cluster is completely depending on the initial value of the coupled
nodes which make it difficult to predict the jump point. Therefore,
$N_s$ can be the number of nodes in the red area ($N_r$) in Fig.
\ref{fig4}(d) or become N all at a sudden for any coupling strength
$\epsilon$ between $[0.75,1.0)$. We can only predict the boundary
value of R before the buffer nodes jumped to synchronous in a sudden
by setting $N_s=N_r$ as shown the dashed lines in Fig.
\ref{fig5}(a). The numerical evidences shown that the jump point are
related to the initial value of the coupled phase oscillators as
shown in Fig. \ref{fig5}(b) for $N=100$ coupled oscillators.
Moreover, the jump point is also related to the system size
according to plot of the order parameter versus $\epsilon$ for
$N=100,500,1000$ as shown in Fig. \ref{fig5}(c). Therefore, the
buffer nodes may join the synchronous cluster in a sudden when the
remaining buffer nodes are less than a certain number $m$ which
results in an earlier jump of the order parameter $R$. Moreover, the
value of $m$ is related to the system size and the initial value of
the coupled oscillators.

\section{Influence of coupling weight on ES dynamics}
 It has been shown that the generalized
Kuramoto model with frequency-weighted coupling can generate
first-order synchronization transition in general networks
\cite{4,25}. The effects of coupling weight on the transition
process in a ring of coupled oscillators are non-trivial. It is
convinient to control the transition process by introducing proper
coupling weight. To reveal the influence of the weights on the
transition process, coupling weights are introduced into the
Kuramoto model similar as that in the complex network\cite{4}.
\begin{eqnarray}\label{eq15}
\dot{\theta_i}(t)=\omega_{i}+\frac{\epsilon}{k_i+1}\sum_{j=1}^{N}
A_{i,j} |{\omega_j-\omega_i}| ^{\alpha}
\sin(\theta_j-\theta_i),i=1,2...,N,
\end{eqnarray}
\begin{figure}
\includegraphics[width=16cm]{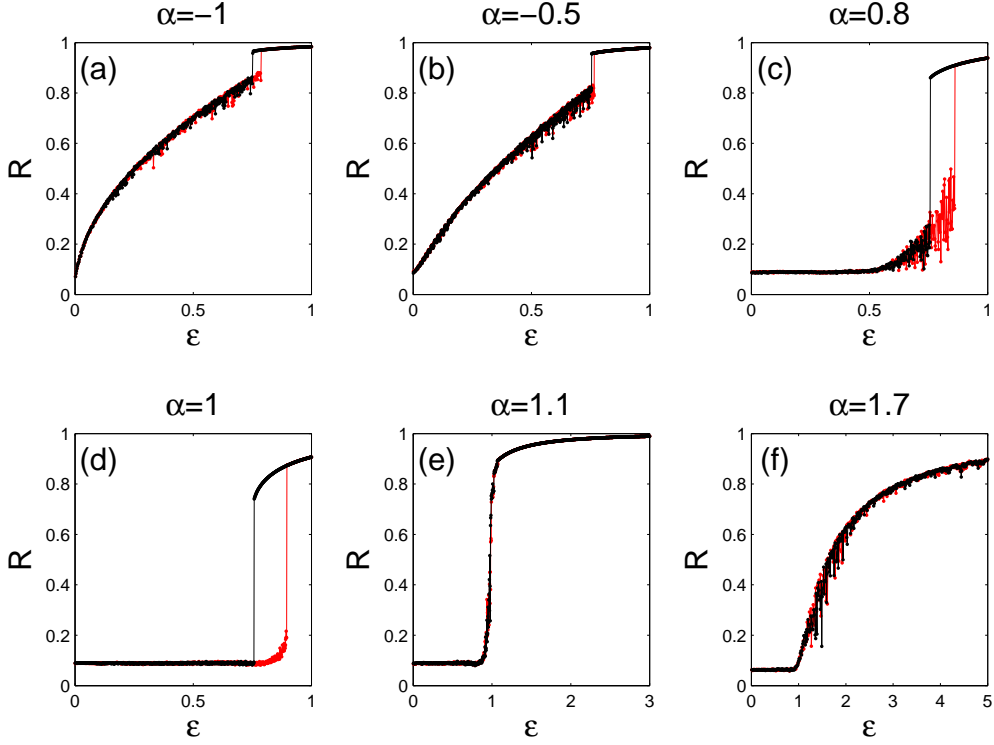}
\caption{(color online) (a)-(f) The numerical results of the order
parameter for the forward (red) and backward (black) traversal at
different coupling weight $\alpha=-1.0,-0.5,0.8,1.0,1.1,1.7$,
respectively.} \label{fig6}
\end{figure}
\begin{figure}
\includegraphics[width=16cm]{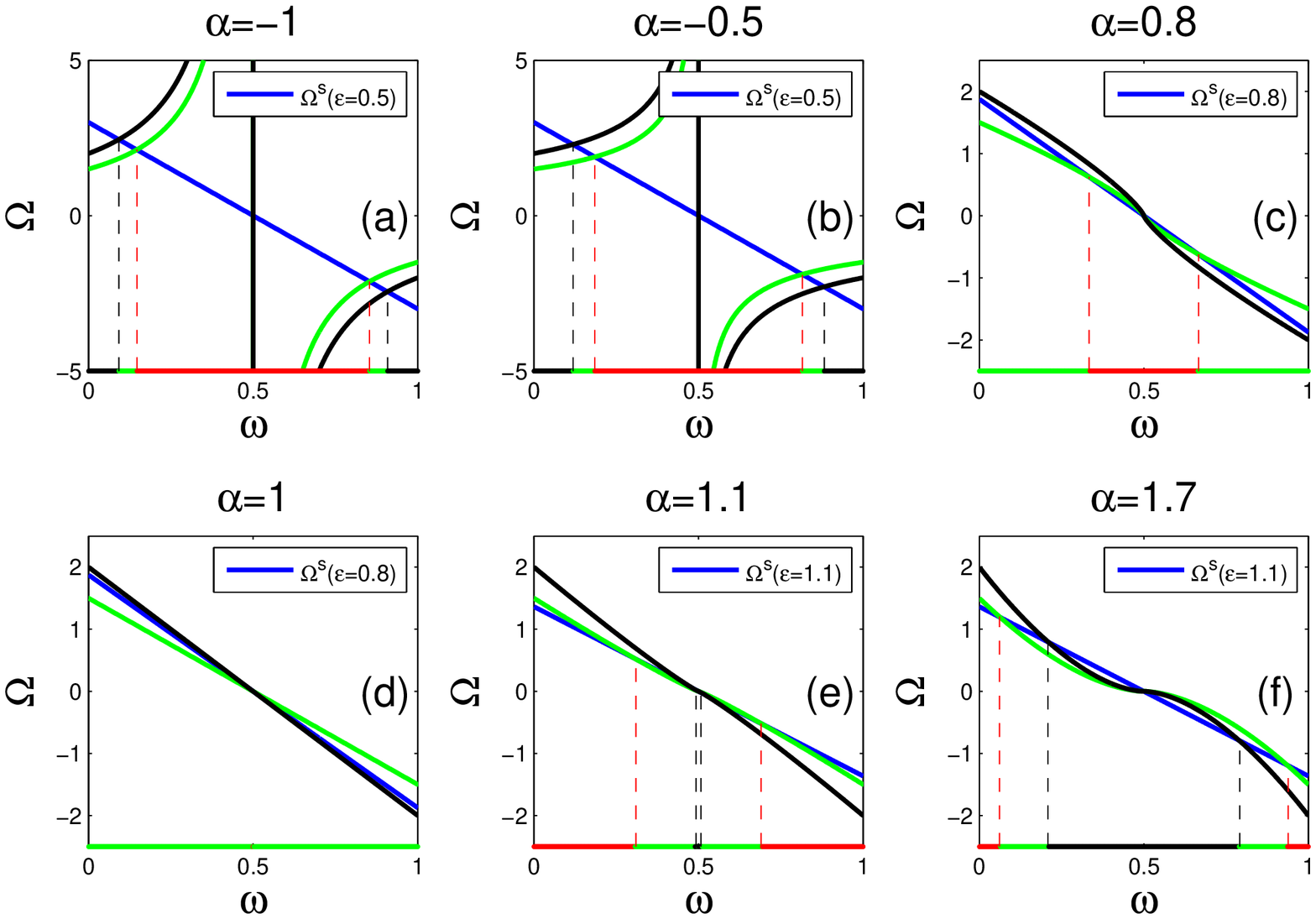}
\caption{(color online) (a)-(f) The boundary lines of the whole
synchronization  $\Omega_{bws}$ (black lines) and partial
synchronization $\Omega_{bps}$ (green lines), and the curves of
$\Omega_i$ (blue lines) versus natural frequency $\omega_i$ for
different coupling weight coefficients
$\alpha=-1.0,-0.5,0.8,1.0,1.1,1.7$.} \label{fig7}
\end{figure}

where $A_{i,j}$ is the same as that in Eq. \ref{eq1}. By increasing
the value of $\alpha$ from 0 to 1 or decreasing it from 0 to -1, the
transition process looks quite different.  In the latter case, the
mixed ES still exists while the boundary curve of the  transition
region changes from linear to convex, leading to the shrinkage of
the area of the bistable part. However, in the former case, the
boundary curve becomes concave, resulting in an expansion of the
area. When $\alpha=1$, the mixed ES turns into a pure ES. However,
it becomes continuous for $\alpha$ slightly larger than $1$. The
dependence on $\alpha$ of the transition process is displayed in
Figs. \ref{fig6}(a)-(f) for $\alpha=-1,-0.5,0.8,1.0,1.1,1.7$.

The coupling weight exponent $\alpha$ will change the number of
nodes $N_s$ in the synchronous cluster as shown in Fig. \ref{fig6}.
The $|\Omega_{i}|$ at the full synchronization for arbitrarily given
$\alpha$ can be written as
\begin{eqnarray}\label{eq16}
\Omega_{bws}=(|\omega_{i+1}-\omega_i|^\alpha
+|\omega_{i-1}-\omega_i|^\alpha),
\end{eqnarray}
while the transition curve to the partial synchronization is
determined by
\begin{eqnarray}\label{eq17}
\Omega_{bps}=0.75(|\omega_{i+1}-\omega_i|^\alpha
+|\omega_{i-1}-\omega_i|^\alpha).
\end{eqnarray}

\begin{figure}
\includegraphics[width=16cm]{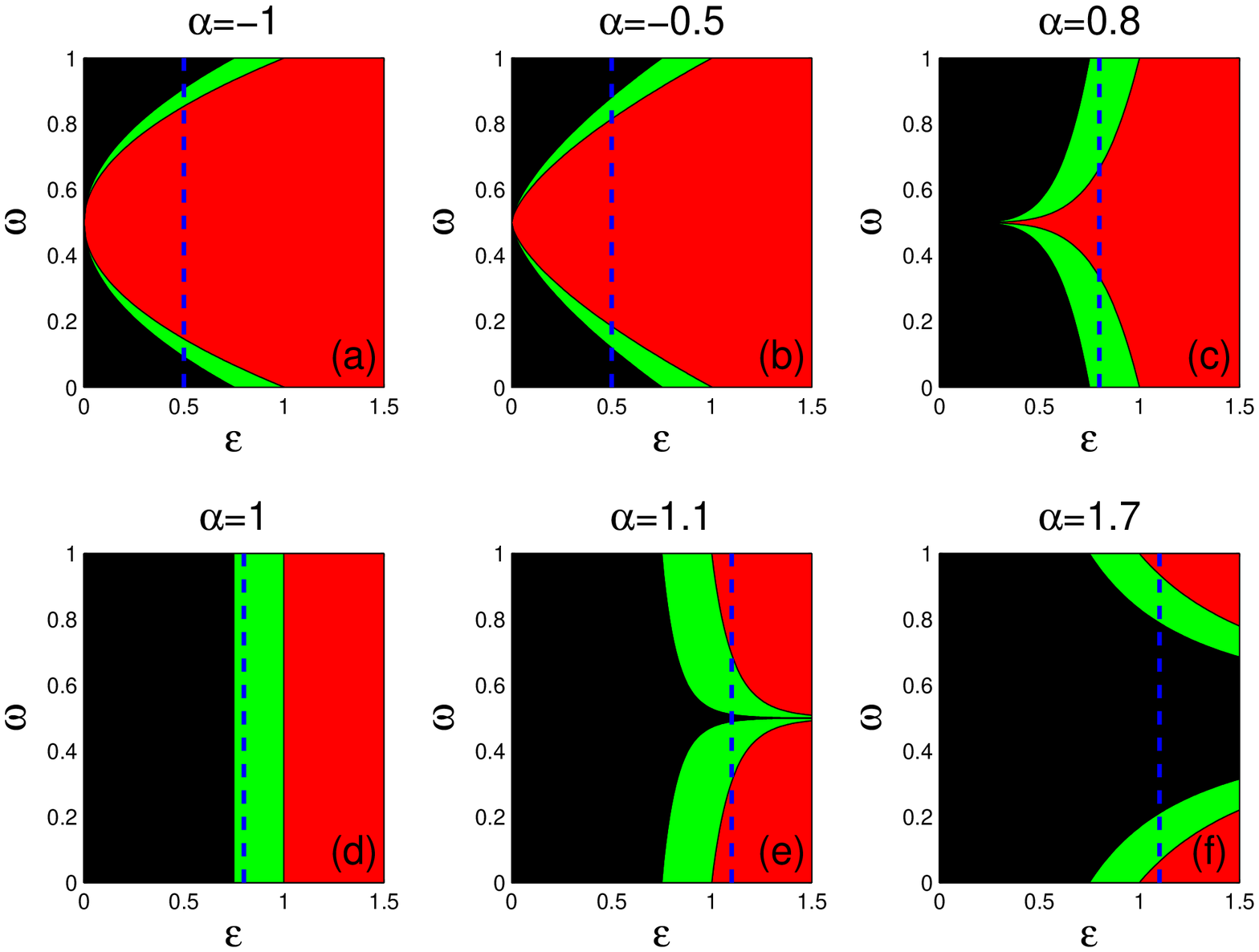}
\caption{(color online) (a)-(f) Three types of states as: the
non-synchronous (black), the buffer (green), and the synchronous
state(red) in the parameter space of $\omega$ versus $\epsilon$ for
different coupling weight exponents
$\alpha=-1.0,-0.5,0.8,1.0,1.1,1.7$.} \label{fig8}
\end{figure}
\begin{figure}
\includegraphics[width=16cm]{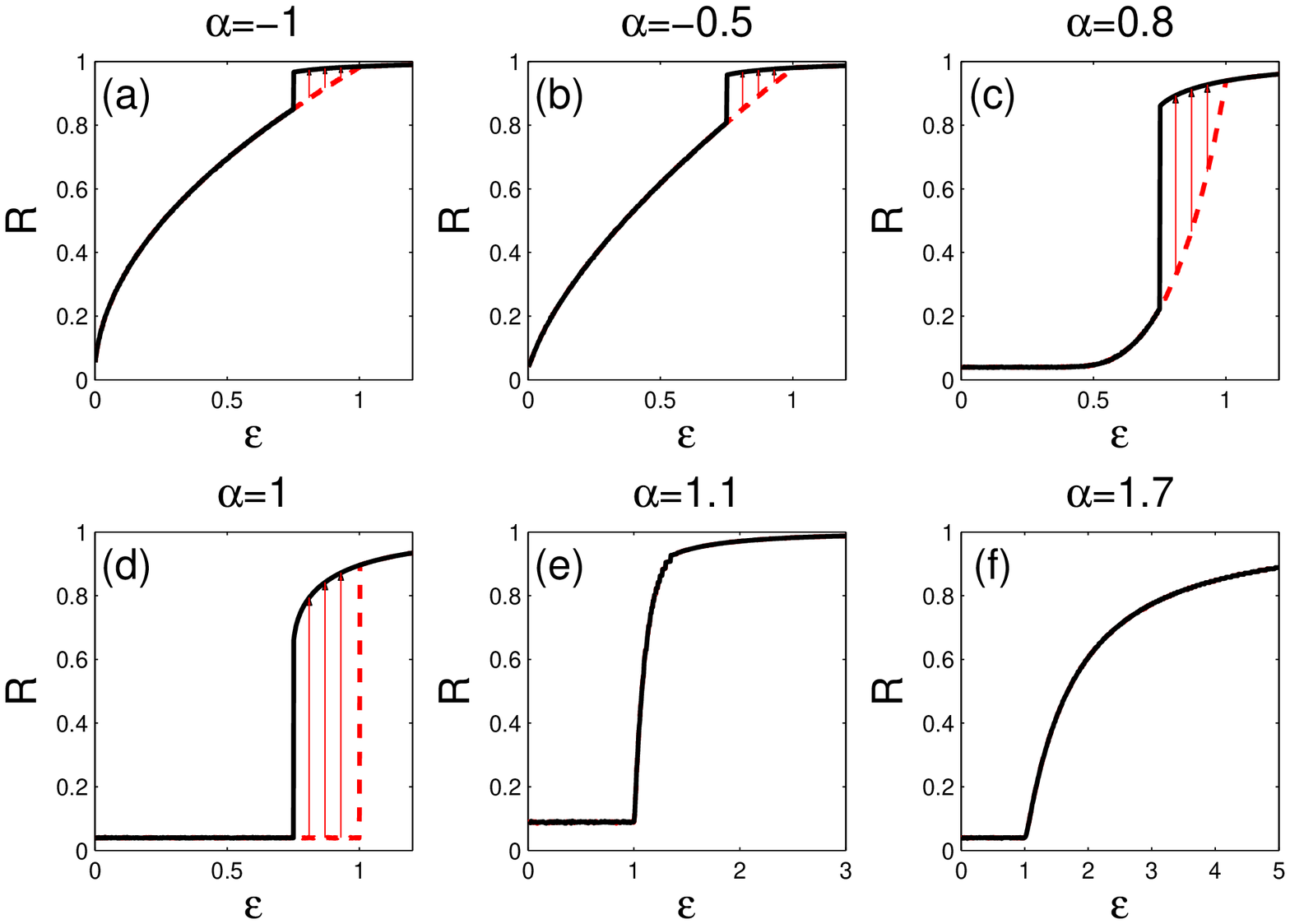}
\caption{(color online) (a)-(f) The order parameters $R$
in the forward (red) and backward (black) direction versus $\epsilon$ for different coupling weight exponents $\alpha=-1.0,-0.5,0.8,1.0,1.1,1.7$.}\label{fig9}
\end{figure}

Therefore, the coupling weight exponent $\alpha$ changes the shapes
of boundary line $\Omega_{bws}$ and $\Omega_{bps}$  as shown in
Figs. \ref{fig7}(a)-(f) for $\alpha=-1,-0.5,0.8,1.0,1.1,1.7$. Two
horizontal lines of $\Omega_{bws}$ and $\Omega_{bps}$ at $\alpha=0$
become hyperbolic for negative $\alpha$, convex for $\alpha \in
(0,1)$, and concave for $\alpha>1.0$. According to Eq. \ref{eq6},
the coupling strength $\epsilon$ determines the slope of $\Omega_i$
for each $\omega_i$. The synchronous nodes (red dots),buffer
nodes(green dots), and the non-synchronous nodes (black dots) are
classified by the two boundary lines and the slope of the line
$\Omega_i$ as shown in Figs. \ref{fig7}(a)-(f). Those with frequency
near $\bar{\omega}$ will get to the synchronous cluster quickly if
$\alpha<1$. On the contrary, those with frequency away from
$\bar{\omega}$ will get to the synchronous cluster first if
$\alpha>1$. Nevertheless, if  $\alpha=1$, all nodes get to
synchronization simultaneously which leads to ES.

The dependence on $\alpha$ of the synchronous process in the
parameter space of $\omega\sim \epsilon$ for
$\alpha=-1,-0.5,0.8,1.0,1.1,1.7$ is shown in Figs.
\ref{fig8}(a)-(f). The red, green, and black area denotes the
synchronous, the buffer, and the non-synchronous state. The red area
increases (decreases) by changing its shape from triangle to bell
(cone) when $\alpha$ varies from $0$ to some negative (positive)
values. When $\alpha=1$, the system makes a transition from the
non-synchronous to the bistable and finally to the synchronous state
as the coupling strength $\epsilon$ increases from $0$ to $1$, which
finally leads to the ES. When $\alpha$ is slightly larger than $1$
(for example $\alpha=1.1$), the square area associated with the
synchronous state is invaded by the cone-shaped buffer and the
non-synchronous state. The number $N_s$ of the synchronized nodes
can be extracted from Figs. \ref{fig8}(a)-(f) for given $\alpha$ and
$\epsilon$. Based on Eqs. \ref{eq13},\ref{eq14}, the order parameter
$R$ for $\alpha=-1.0,-0.5,0.8,1.0,1.1,1.7$ are calculated with the
same method as those in Fig. \ref{fig5}(a) and shown in Figs.
\ref{fig9}(a)-(f). The theoretical results again match well with the
numerics.

\section{Discussion and conclusion}
In conclusion, we have firstly observed a kind of mixed ES dynamics
in the ring of a coupled phase oscillators by introducing a special
frequency configuration of nodes. With this configuration, the
coupled oscillators transit from partial synchronization to ES with
the increment of the coupling strength. Generally, the coupling
strength required for the synchronization of two coupled oscillators
increases with the increment of their frequency mismatch. Therefore,
those nodes whose frequencies are near the average get to
synchronization first and form a synchronous cluster. The mixed ES
dynamics can be well understood by the competition between three
types of nodes (non-synchronous, buffer and synchronized nodes) in
the nearest-neighbor coupled phase oscillators. In the partial
synchronization process, the interaction between the non-synchronous
and the synchronous nodes is transmitted by the buffer nodes, so
that the increasing coupling strength tends to increase the number
of the synchronous nodes while decreasing the non-synchronous ones.
Once the non-synchronous nodes diminished, the competition exists
only between the synchronous and the buffer nodes, so that the
buffer nodes may stay non-synchronous or join into the synchronous
cluster in a sudden depending on their initial condition and system
size. As a result, the system transits dynamics change from partial
synchronization to ES.

Upon introducing the coupling weight, the dynamics of the coupled
phase oscillators are determined by the competition between the
frequency mismatch and the coupling weight. With the negative
coupling weight exponent $\alpha$, the larger frequency mismatch,
the more weakly coupled the two nodes are. Since the negative
coupling weight exponent $\alpha$ increases the effective coupling,
the synchronous (non-synchronous) nodes increase (decrease) for a
given coupling strength $\epsilon$, which leads to a larger order
parameter. On the contrary, the positive coupling weight exponent
$\alpha$ acts oppositely, which leads to a smaller order parameter.
As the coupling weight exponent $\alpha=1$, the effective coupling
strength is proportional to the frequency mismatch, which makes all
nodes have the same critical coupling strength to get
synchronization. Therefore, we observe the ES but not the partial
synchronization. As $\alpha>1$, the nodes with larger frequency
mismatch have larger coupling strength which again makes it easier
to get synchronization than those with smaller frequency mismatch.
The ES dynamics is thus not possible and replaced by a continuous
transition.

Our results are based on the special frequency distribution. In
order to realize ES or mixed ES, it is necessary to avoid
multi-synchronous clusters. Therefore, the frequency distribution
must be strictly symmetric with respect to the average frequency of
the whole coupled system. Moreover, the nodes whose frequencies are
near the average frequency must be spatially near each other. The
frequencies of nodes can be randomly selected so long as they
fulfill the two conditions above.  Our results may help understand
the synchronization in more complex networks with homogenous
coupling and shed light on the control of pattern formation dynamics
of coupled oscillators\cite{32,33,34}.

%\begin{figure}
%\includegraphics[width=8cm]{fig10.eps}
%\caption{(color online)The areas of hysteresis versus the coupling weight coefficients $\alpha$.}\label{fig10}
%\end{figure}

%\section{Acknowledgments}

\begin{acknowledgments}
This work is supported by the National Natural Science Foundation of
China (NSFC) (Grants Nos. 61377067 ,11775034, 11375033), Weiqing Liu
is supported by the training plan of young scientists of Jiangxi
Province and the Qingjiang Program for Excellent Young Talents,
Jiangxi University of Science and Technology.
\end{acknowledgments}

%\bigskip \newpage

%\newpage
\end{document}